\def\MYJOURNAL{1} %IEEE

\if 0\MYJOURNAL% 
\documentclass[3p, preprint, sort&compress]{elsarticle}
\usepackage{setspace}
\else \if 1\MYJOURNAL%
\documentclass[9pt,compsoc,letter]{IEEEtran}
\usepackage{geometry}%
\usepackage[numbers,sort&compress]{natbib}
\usepackage{setspace}

\fi
\fi

\usepackage{color, xcolor, colortbl}
\usepackage{hyperref}
\usepackage{amsfonts}
\usepackage{amsmath}
\usepackage{amssymb}
\usepackage{graphicx}
\usepackage{url}
\usepackage{fp}
\providecommand{\U}[1]{\protect\rule{.1in}{.1in}}

\usepackage{enumitem}
\usepackage[referable,para]{threeparttablex}
\usepackage{footnote}
\usepackage{booktabs}  % for \toprule ...

\usepackage{mwe}
\usepackage[percent]{overpic}

\usepackage{listings}
\definecolor{dkgreen}{rgb}{0,.6,0}
\definecolor{dkblue}{rgb}{0,0,.6}
\definecolor{dkyellow}{cmyk}{0,0,.8,.3}

\lstdefinestyle{customphp}{
  language        = php,
  breaklines=true, breakindent=38pt,
  keywordstyle    = \color{dkblue},
  stringstyle     = \color{red},
  identifierstyle = \color{dkgreen},
  commentstyle    = \color{gray},
  emph            =[1]{php,ELEMENT,GUIDE},
  emphstyle       =[1]\color{black},
  emph            =[2]{if,and,or,else},
  emphstyle       =[2]\color{dkyellow},
  basicstyle      = \scriptsize\ttfamily}

\lstdefinestyle{customc}{
  breaklines=true, breakindent=20pt,
  frame=leftline,
  numbers=left,
  language=C, numberstyle=\tiny, numbersep=10pt,
  showstringspaces=false,
  basicstyle=\footnotesize\ttfamily,
  keywordstyle=\bfseries\color{green!40!black},
  commentstyle=\itshape\color{purple!40!black},
  identifierstyle=\color{blue},
  stringstyle=\color{orange},
  captionpos=t
}

\hypersetup{
  colorlinks,
  citecolor=blue,
  filecolor=blue,
  linkcolor=blue,
  urlcolor=blue
}

\graphicspath{{./Graphics/}}

\usepackage{fourier}
\DeclareMathAlphabet{\mathcal}{OT1}{pzc}{m}{it}
\DeclareSymbolFont{letters}{OML}{cmm}{m}{it}

\usepackage{tikz}
\usepackage{float}
\usepackage{pgf}
\usepackage{pgfplots}

\usetikzlibrary{calc}
\usepgflibrary{shapes}
\usetikzlibrary{er}
\usetikzlibrary{arrows,snakes,backgrounds}
\usetikzlibrary{decorations.text}

\usepgfplotslibrary{external}
\usetikzlibrary{spy}

\def\getangle(#1) (#2)#3{%
  \begingroup%
  \pgftransformreset%
  \pgfmathanglebetweenpoints{\pgfpointanchor{#1}{center}}{\pgfpointanchor{#2}{center}}%
  \expandafter\xdef\csname angle#3\endcsname{\pgfmathresult}%
  \endgroup%
}

\pgfplotsset{compat=1.11}
\usetikzlibrary{calc,trees,positioning,arrows,chains,shapes.geometric,%
  decorations.pathreplacing,decorations.pathmorphing,shapes,%
  matrix,shapes.symbols}

\tikzset{
  >=stealth',
  punktchain/.style={
    font=\scriptsize,
    rectangle,
    rounded corners,
    draw=black, thick,
    text width=10em,
    minimum height=1em,
    text centered},
  line/.style={draw, thick, <-},
  element/.style={
    tape,
    top color=white,
    bottom color=blue!50!black!60!,
    minimum width=8em,
    draw=blue!40!black!90, very thick,
    text width=10em,
    minimum height=1em,
    text centered},
  every join/.style={->, thick,shorten >=1pt},
  decoration={brace},
  tuborg/.style={decorate},
  tubnode/.style={midway, right=2pt},
}

\if 0\MYJOURNAL%
\tikzset{
  PIXEL/.style={
    font=\fontsize{4}{3.6}\selectfont,
    text width=9em,
    minimum height=1em,
    text centered
  }
}
\else \if 1\MYJOURNAL%
\tikzset{
  PIXEL/.style={
    font=\tiny,
    text width=8em,
    minimum height=3em,
    text centered
  }
}
\fi
\fi

\usepackage{multirow}
\usepackage{subfigure}

\definecolor{mygreen}{rgb}{0,0.6,0}
\definecolor{mygray}{rgb}{0.5,0.5,0.5}
\definecolor{mymauve}{rgb}{0.58,0,0.82}

\definecolor{darkgray}{rgb}{.4,.4,.4}
\definecolor{purple}{rgb}{0.65, 0.12, 0.82}

\lstdefinelanguage{JavaScript}{
  keywordstyle=\color{blue}\bfseries,
  ndkeywords={class, export, boolean, throw, implements, import, this},
  ndkeywordstyle=\color{darkgray}\bfseries,
  identifierstyle=\color{black},
  sensitive=false,
  comment=[l]{//},
  morecomment=[s]{/*}{*/},
  commentstyle=\color{purple}\ttfamily,
  stringstyle=\color{black}\ttfamily,
  morestring=[b]',
  morestring=[b]"
}

\lstdefinestyle{customJS}{
  language=JavaScript,
  extendedchars=true,
  basicstyle=\footnotesize\ttfamily,
  showstringspaces=false,
  showspaces=false,
  numbers=left,
  numberstyle=\tiny,
  numbersep=0pt,
  tabsize=2,
  breaklines=true,
  showtabs=false,
  captionpos=b
}

\usepackage{multicol}
\usepackage{lipsum}                     % Dummytext
\usepackage{xargs}                      % Use more than one optional parameter in a new commands
\usepackage{xcolor}  % Coloured text etc.
\usepackage{filecontents}

\usepackage[colorinlistoftodos,prependcaption,textsize=tiny]{todonotes}
\newcommandx{\unsure}[2][1=]{\todo[linecolor=red,backgroundcolor=red!25,bordercolor=red,#1]{#2}}
\newcommandx{\change}[2][1=]{\todo[linecolor=blue,backgroundcolor=blue!25,bordercolor=blue,#1]{#2}}
\newcommandx{\info}[2][1=]{\todo[linecolor=OliveGreen,backgroundcolor=OliveGreen!25,bordercolor=OliveGreen,#1]{#2}}
\newcommandx{\improvement}[2][1=]{\todo[linecolor=red,backgroundcolor=red!25,bordercolor=red,#1]{#2}}
\newcommandx{\thiswillnotshow}[2][1=]{\todo[disable,#1]{#2}}

\begin{document}

%\onecolumn
\title{DVP: Data Visualization Platform}

\author{%
  Waleed A. Yousef\textsuperscript{a,b},~\IEEEmembership{Senior Member,~IEEE,}~\thanks{Yousef, Waleed
    A., is an associate professor, \url{wyousef@GWU.edu}}

  Ahmed A. Abouelkahire\textsuperscript{c},~\thanks{Ahmed A. Abouelkahire, B.Sc., Senior Data
    Scientist, TeraData, Egypt, \url{ahmedanis03@gmail.com}}
  Omar~S.~Marzouk\textsuperscript{c},~\thanks{Omar S. Marzouk, M.Sc., School of Computer Science and
    Communication (CSC), KTH Royal Institute of Technology, Sweden, \url{omares@kth.se}}

  Sameh~K.\ Mohamed,~\thanks{Sameh K. Mohamed, M.Sc., Insight Center for Data Analytics, National
    University of Ireland, Ireland, \url{sameh.kamal@insight-centre.org}}

  Mohammad Alaggan\textsuperscript{a,b},~\thanks{Mohammad Alaggan, is an assistant professor,
    \url{MAlaggan@fci.Helwan.edu.eg}}

  \begin{center}
    \hfil\includegraphics[height=0.21\textheight]{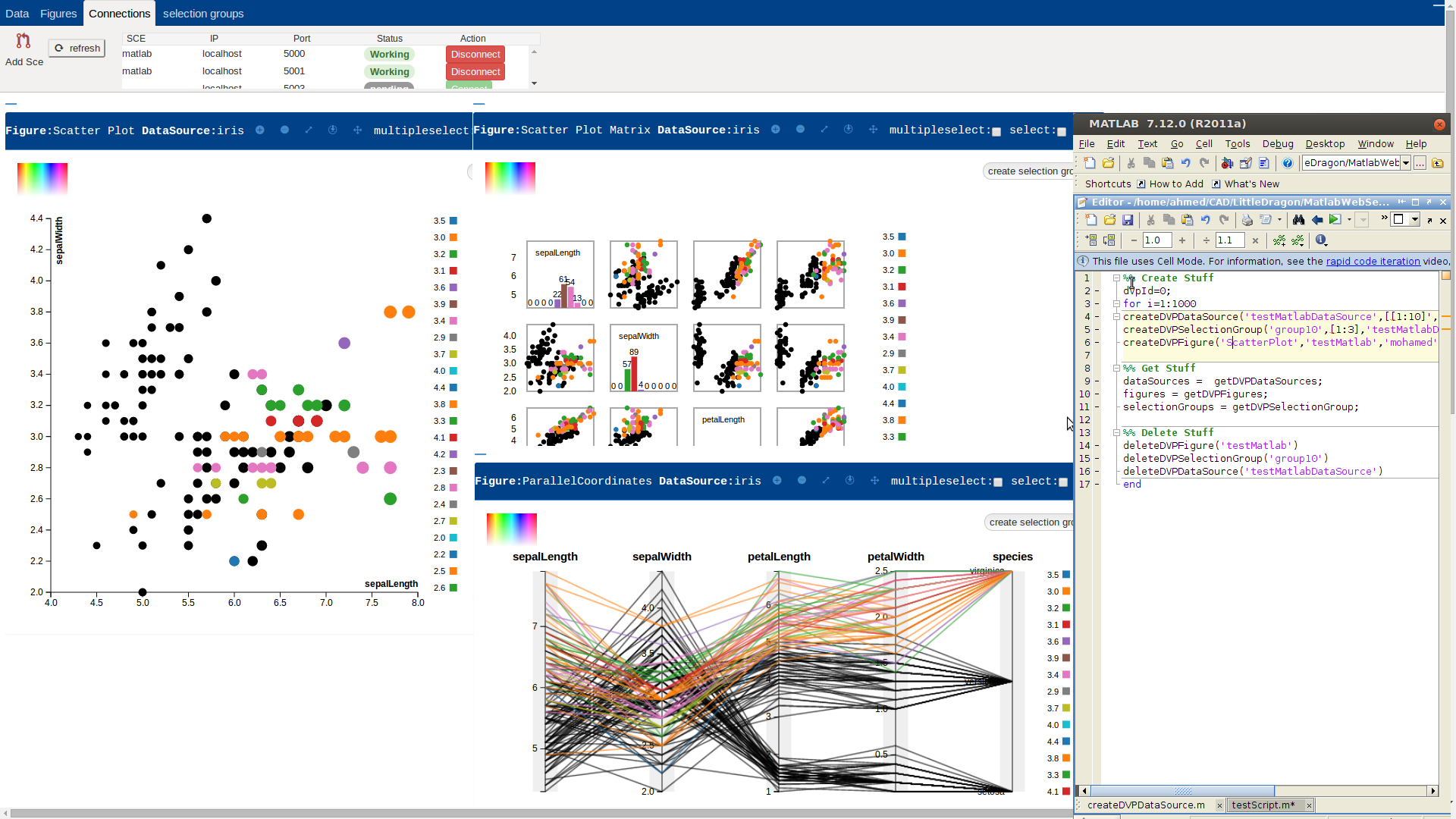}\hfil\includegraphics[height=0.21\textheight]{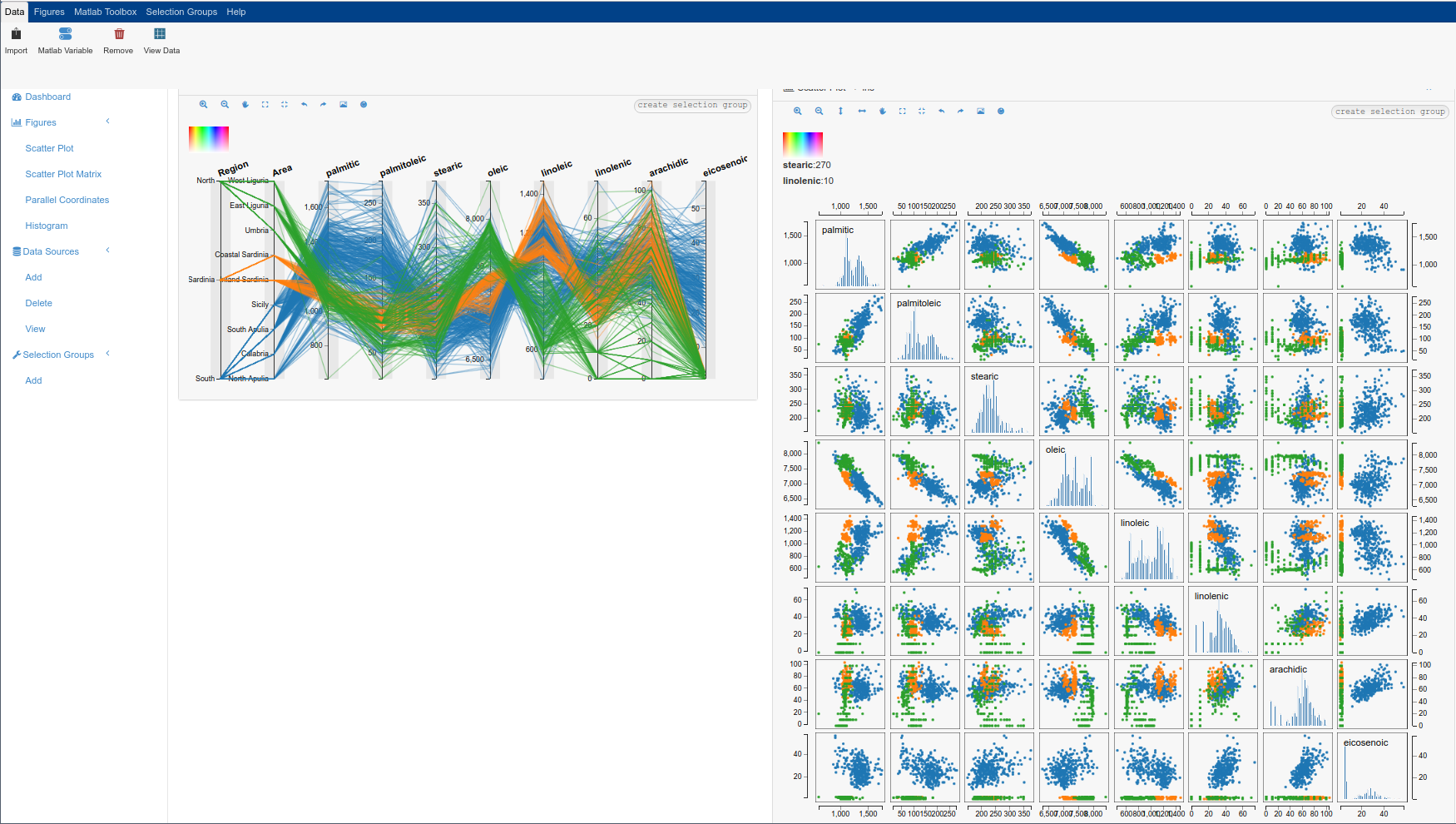}\hfil

    {\footnotesize\ Two snapshots for DVP. Left: the DVP integrated locally to Matlab on the
      desktop and acts as a Matlab toolbox, where all variables are communicated from/to
      Matlab. Right: the online version of the DVP where all actions, interactions, and dynamics can
      be performed.}
  \end{center}

  \thanks{\textsuperscript{a}Human Computer Interaction Laboratory (HCI Lab.), \url{http://hciegypt.com/}, Egypt.}
  \thanks{\textsuperscript{b}Computer Science Department, Faculty of Computers and
    Information, Helwan University, Egypt}
  \thanks{\textsuperscript{c}These authors contributed equally to the manuscript as the second author and names are
    ordered alphabetically according to the family name.}
  \thanks{This project was funded by: MESC Labs for Research and Development \url{www.mesclabs.com}}}

\maketitle

  \begin{abstract}
    We identify two major steps in data analysis, data exploration for understanding and observing
patterns/relationships in data; and construction, design and assessment of various models to
formalize these relationships. For each step, there exists a large set of tools and software. For
the first step, many visualization tools exist, such as, GGobi, Parallax, and Crystal Vision, and
most recently tableau and plottly. For the second step, many Scientific Computing Environments
(SCEs) exist, such as, Matlab, Mathematica, R and Python. However, there does not exist a tool which
allows for seamless two-way interaction between visualization tools and SCEs. We have designed and
implemented a data visualization platform (DVP) with an architecture and design that attempts to
bridge this gap. DVP connects seamlessly to SCEs to bring the computational capabilities to the
visualization methods in a single coherent platform. DVP is designed with two interfaces, the
desktop stand alone version and the online interface. To illustrate the power of DVP design, a free
demo for the online interface of DVP is available \citep{DVP} and very low-level design details are
explained in this article. Since DVP was launched, circa 2012, the present manuscript was not
published since today for commercialization and patent considerations.
  \end{abstract}

  \newcommand{\sep}{,~} %to mach that of PR

%%% Local Variables:
%%% mode: latex
%%% TeX-master: "Yousef2016DVP003"
%%% End:

\section{Introduction}\label{SecIntro}
\subsection{Why Data Visualization?}\label{sec:need-data-visu}
Data acquisition is ubiquitous; and data arise from diverse areas and applications, including
medical, financial, industrial, governmental, among others. Data of size $n\times p$ consist of $n$
records/observations, and each consists of $p$ dimensions/features. When $p$ increases dramatically
data is high dimensional. An example is DNA microarray data where the number of observations (here
are patients) is in the order of hundreds, while the number of dimensions (here are genes) is in the
order of thousands. When $n$ increases dramatically data is called ``big''. An example is
astronomical data where the number of observations reaches billions!

Regardless to the origin of data or its application---data analysis, including statistics,
statistical learning, machine learning, and pattern recognition, collectively are concerned with
understanding data, recognizing patterns, and learning input-output relationships hiding in
data. Modeling such a pattern/relationship can be described by a regression function, classification
rule, clustering analysis, or mere statistical testing and summaries. This modeling is used for
prediction (or decision support), and interpretation. Two steps usually are involved:

\begin{enumerate}[partopsep=0.05in,parsep=0.05in,topsep=0.05in,itemsep=0.05in,leftmargin=0.1in]
  \item Data exploration and visualization for understanding and observing patterns/relationships in
  data. This step involves visualizing data in many interactive and dynamic plots. Each plot conveys
  part of the story, which is emphasized by the interaction with each plot and linking among
  different plots; \citep[see,
  e.g.,][]{Chen2008DataVisualization,Wegman2003VisualDataMining,Wegman1992TheGrandTour,Inselberg2011TutorialDataVisualization,Inselberg2000ClassVisu,Inselberg2002VisData}. The
  term ``Data Visualization'' is used interchangeably with ``Exploratory Data Analysis'' (EDA) and
  recently more fashionably ``Visual Data Mining''; all convey the meaning and objective behind such
  a step.

  \item Construction, design, and assessment of the model to formalize these patterns/relationships
  and to assess their statistical significance and generalization to the population of data;
  \citep[see,
  e.g.,][]{Cherkassky1998LearningFrom,Fukunaga1990Introduction,Hastie2001TheElements,Bishop2006PatRecMachInt,Vapnik2000Nature,Vapnik1998StatLerningTh}.
\end{enumerate}

Each of these two steps accounts for a field by itself with its own literature, theory, and
software. Although inclusion of the two steps results in a consolidated design and great
understanding of data, not all practitioners adopt such a comprehensive view when analyzing
data. The need for the first step, data visualization and exploration, becomes more crucial when
data become high dimensional (huge $p$) or become ``big'' (huge $n$). This is true since modeling,
analyzing, and processing data with huge $n$ or/and $p$ become more difficult and complicated. Data
visualization and exploration reveal secrets and paves shortcuts to understanding data and building
best models.

\subsection{Why DVP for Data Visualization?}\label{sec:why-dvp-as}

%\begin{figure}\centering
%  \includegraphics[width=0.5\textwidth]{./Graphics/FigComparingDVP.png}
%  \caption{DVP vs. other Data Visualization Software (DVSW) based on 2013 versions of all of
%    them.}\label{FigComparingDVP}
%\end{figure}

\begin{table*}[t]
	\centering
	\caption{Comparison of popular visualization software packages. 'x' denotes full support and '-' denotes partial support.}
	\label{tab:dvp-tools}
	\resizebox{.999\textwidth}{!}{
	\begin{tabular}{ccccccccc}
		\toprule
		Package & SCE support & Interactivity & Linked Plots & Extensibility & Cross Platform & Figure Customisation & Database Servers & Mutliple devices \\
		\midrule
		Matlab      & -           & -             &              & -             & x              & -                    & x                &                  \\
		Mathematica & -           & -             & -            &               & x              &                      & x                & -                \\
		SigmaPlot   &             & -             &              &               &                & -                    &                  &                  \\
		OriginLab   &             &               &              & -             &                & -                    &                  &                  \\
		Tablue      &             & -             & -            & -             & -              & -                    & x                & x                \\
		PV-Wave     &             & x             & -            & -             & -              &                      & x                &                  \\
		Plot.ly     & -           & -             & -            & -             & -              &                      &                  & x                \\
		Parallex    & -           & x             & -            & -             &                &                      & -                &                  \\
		RGgobie     & -           & x             & -            & -             &                &                      &                  &                  \\
		DVP         & x           & x             & x            & x             & x              & x                    & x                & -               \\
		\hline
	\end{tabular}
	}
	
\end{table*}

Many data visualization software (DVSW) exist that can produce similar results with the capabilities
of ``interaction'' and ``linking'', which are not supported by any Scientific Computing Environment
(SEC) as Matlab, Mathematica, SAS, etc. Then, the right question is this: what is the need for
another DVSW, and why do we propose our Data Visualization Platform (DVP)? We provide below, in
bullets, an answer for this question and show how DVP design and philosophy is important to
scientists, researchers, and data analysts in different fields. Although not all of the following
aspects are currently implemented, but the DVP kernel is designed with eyes on the following:

\begin{itemize}[partopsep=0.1in,parsep=0.05in,topsep=0.05in,itemsep=0.05in,leftmargin=0.1in]
  \item\textbf{Seamless communication with any SCE to behave as a single environment.} Current DVSWs
  are standalone software that are detached from SCEs. Any scientist, researcher, or data analyst
  using any SCE cannot interact with patterns visualized and discovered in the DVSW. For example, if
  the data analyst uses Matlab to analyze a dataset, and Parallax to visualize data, he cannot do
  processing on patterns discovered in Parallax; these patterns are not, of course, seen as
  variables in Matlab workspace. It is impossible to iterate back and forth between the DVSW and the
  SCE except by tedious data export and import that puts hurdles. DVP is designed to interact
  seamlessly with any SCE, as if both are one environment, even if DVP and SCE are running on two
  different machines. This is extremely important for connecting with computing clouds for analyzing
  big data.

  \item\textbf{High extensibility to different plots and methods in various scientific fields.}
  Current DVSWs provide some visualization methods, e.g., ||-coords, scatter plot, matrix plot,
  projection pursuit, grand tour in 4 dimensions, etc. However, many scientific fields require more
  sophisticated methods. For example, graph analysis and astronomical data require Multi-Dimensional
  Scaling (MDS) plots \citep[see, e.g.,][II.6]{Chen2008DataVisualization}. It is almost impossible
  for any DVSW to provide all the available plotting and charting methods, let alone ones being
  continuously developed from many fields of science. DVP, in addition to the wide range of plots it
  provides, it is designed to provide an easy scripting language based on JavaScript that enables
  users to write their own plotting methods and integrate them to DVP. This will build a wider user
  community and enrich it with many sophisticated methods.

  \item\textbf{Support data from network streams and common local database servers, e.g., SQL,
    MySQL, and Oracle.} Many DVSWs only load data statically from a local machine storage. However,
  nowadays, many data sources belonging to many applications are available online and updated in
  real time; e.g., stock market data, data of global enterprises, Yahoo data, etc. Analysts
  monitoring such data have to be connected all the time. DVP is designed to facilitate connection
  to network streaming and different online database sources.

  \item\textbf{Available API for interfacing with different hardware, e.g., Raspberry Pi and Arduino
    chips.} Data acquisition is not explicit to software and reports; data are acquired from
  hardware as well, e.g., Arduino and Raspberry Pi chips. Arduino \citep{arduino} is a micro
  controller designed with the objective to connect to the ambient; a chip has different sensors for
  humidity, light, and moisture, etc. Raspberry Pi \citep{raspberry} is a credit-card size computer;
  yet it is so simple that anyone can program it. DVP is designed to provide API for interfacing
  with hardware devices.

  \item\textbf{Cross platform compatibility, e.g., Windows, Linux, Mac, and iOS.} In contrast to
  many available DVSWs, DVP is designed to operate across different operating systems.

  \item\textbf{Multi-device rendering support, e.g., touch screens, big data displays, dashboards,
    and interactive PDFs} Many DVSWs render only to desktop screen, they are not designed for
  displaying big data on large displays, although there is a demand to render data to large displays
  as we have entered already the era of big data. DVP is designed to render to small and large
  displays and to receive input from touch devices as well for wider user community and commercial
  needs. DVP is designed to offer business solutions, as well, for enterprises by supporting
  web-based dashboards and online visualization. In addition, DVP is designed to produce interactive
  PDF documents by exporting figures and plots to PDF with the capability of interacting with those
  figures in the PDF document itself. This integrates reporting schemes to interactive graphics for
  portability and wider utility.

  \item\textbf{Customizable figures and plots.} As opposed to many DVSWs, DVP is designed to provide
  full customization to its figures and plots. Moreover, the design concept behind DVP is that every
  activity is a result of a function call with passed parameters. The GUI actions of the DVP do
  nothing but calling those functions. This means that users can create whatever plots, figures, new
  methods, and fully customize them with the provided scripting language.
\end{itemize}

Table \ref{tab:dvp-tools} is a more quantitative comparison between the first version of DVP and
other well known software available in the market for either data visualization or scientific
computing. The comparison is established on 2013 version of all of them, when DVP was launched. It
is clear that the majority of aspects important to a complete visualization system are missing in
the available systems. DVP is concerned with providing all of these technical features and aspects
in one platform. In Section \ref{SecObjectives}, below, we elaborate more on those features and
detail each of them.

\subsection{Organization of Manuscript}\label{sec:organ-manuscr}
The rest of this paper is organized as follows. Section \ref{SecDetailedResProp} is a high-level
design aspects, requirements, and features of DVP. Sections
\ref{sec:architecture-design}--\ref{chp:SCE-Pluging} detail the design and architecture of DVP
components and subsystems. To clarify the power of the DVP kernel design, we had to present in these
sections some very low-level technical details at the level of variables and processes. Appendix
\ref{sec:backgr-tutor-motiv} is a very short account and tutorial on data visualization and
graphics, taken almost from textbooks, that includes history, importance and motivation for
exploratory data analysis and Grammar of Graphics (GoG). It is important for a reader who is not
fully acquainted with the field; however, it can be trimmed out from the manuscript without
affecting its coherence.

%%% Local Variables:
%%% mode: latex
%%% TeX-master: "Yousef2016DVP003"
%%% End:

\section{DVP: high level design aspects}\label{SecDetailedResProp}
In this section we provide the high level design aspects and philosophy of DVP; some of these
aspects are implemented and others are still under development. Even those aspects that are still
under development are taken care and accounted for the internal design and architecture of the whole
system. For a full account of the implemented aspects the reader may refer to the technical manual
of the DVP.
\subsection{Formal Scripting Language Based on ``Grammar of Graphics''}\label{SecGoG}
In this project of designing a data visualization software, we adopted a very sophisticated design
of scientific plots and figures. It is based on the so called {`Grammar of Graphics'' (GoG) that was
  proposed first by \cite{Wilkinson2006GrammarOfGraphics}, and adopted, e.g., in the R package for
  static plotting \texttt{ggplot2} \citep{Wickham2009ggplot2}. The power of that very formal
  approach of describing plots and figures lies in its generality and in its procedural design. With
  GoG, graphics are not a simple render of colored points on a planar area. Rather, with GoG there
  is a formal language to describe graphics at some level of abstraction. This is so powerful a tool
  for describing new plotting methods or modifying existing ones, which is ideal for our general
  purpose DVP that provides a scripting language for coding new plotting methods. For more
  information on GoG the reader my refer to \cite{Wilkinson2006GrammarOfGraphics}.

\subsection{Algorithms, Mathematics, and Properties of Plots}
There are many important plots for visualizing data that require rigorous mathematical treatment or
algorithm design; e.g., Parallel Coordinates (||-coords), Projection Pursuit (PP), Force-Directed
Graphs (FDG), Multi-Dimensional Scaling (MDS), among others.

One example is ||-coords, where designing useful data group selections (queries) requires good
knowledge of the geometry of high dimensions in ||-coords. Useful mathematical queries have to be
designed for selecting observations of particular interest, e.g., observations with particular
slope, observations with some correlation coefficient, etc. Having such queries in DVP enables us,
e.g., to interact more smartly with the ||-coords plots. For more information on ||-coords the
reader may refer to \cite{Inselberg2008ParCord}.

In addition, algorithm design is needed for some quiries. For example when selecting some data of
interest between two parallel axes in ||-coords. This sounds trivial at the first glance; however,
between the axes one has to search where the GUI selecting tool (e.g., mouse) is moving and
intersecting with drawn lines. Brute-force search is disastrous if not impossible for large
data. Efficient heuristic search algorithms have to be designed, not to sacrifice selection accuracy
for performance optimization and rendering speed.

\subsection{Global System Architecture} \label{SecSolArchitecture}
\begin{figure}[t]
  \centering
  \includegraphics[width=0.5\textwidth]{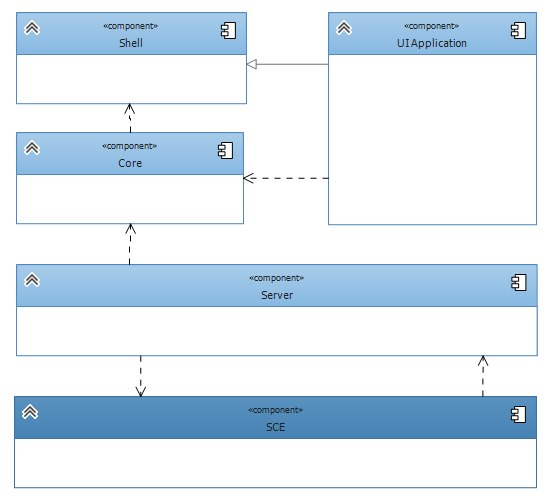}
  \caption{DVP system architecture; (arrows indicate dependency)}\label{FigSysArch}
\end{figure}
DVP is a complex system that provides a various set of innovative features, and these features
depend on a wide set of modules and technologies that construct the main architecture of DVP. As
sketched in Figure \ref{FigSysArch}, the architecture design of the DVP system consists of separate
cooperating parts (subsystems) that work all together to achieve system functionalities. These parts
are responsible for providing the support for cross platform compatibility, multi-device support,
communication with SCEs, and rich interactive visualization. The architecture of each subsystem is
discussed below with a short description of its functionality, structure, and provided features.

\subsubsection{Shell application} \label{sub:shellApp}
\begin{figure}[t]\centering
  \includegraphics[width=0.5\textwidth]{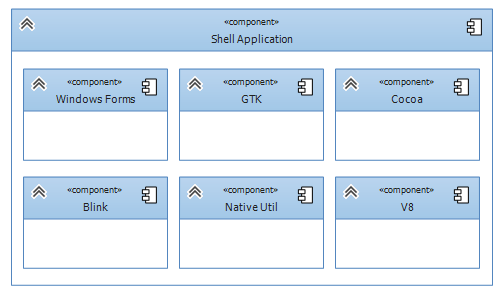}
  \caption{Shell application components}\label{FigShellApp}
\end{figure}
Shell application, as sketched in Figure \ref{FigShellApp}, is a fully functioning web browser that
runs as a container and as a host for web content, which accounts for a perfect way to host HTML
based applications natively on heterogeneous operating systems supported by the browser. This design
can be found in applications like Atom IDE \citep{atom}, Adobe Brackets \citep{brackets}, and
LightTable \citep{lighttable}. This shell application consists of two main subparts on Figure
\ref{FigShellApp}. The first is a platform dependent native UI component, which appears in the
Figure as Windows Forms for Windows, GTK for Linux, and Cocoa for Mac. The second is Chromium
Embedded Framework (CEF), which appears in the Figure as Blink \citep{blink1}, NativeUtil, and V8
\citep{v8}. The first is responsible for cross platform compatibility and the second is responsible
for accessing native resources, linking third part modules, and performing file operations.

\subsubsection{Communication server} \label{SecComServer}
\begin{figure}[t]\centering
  \includegraphics[width=0.5\textwidth]{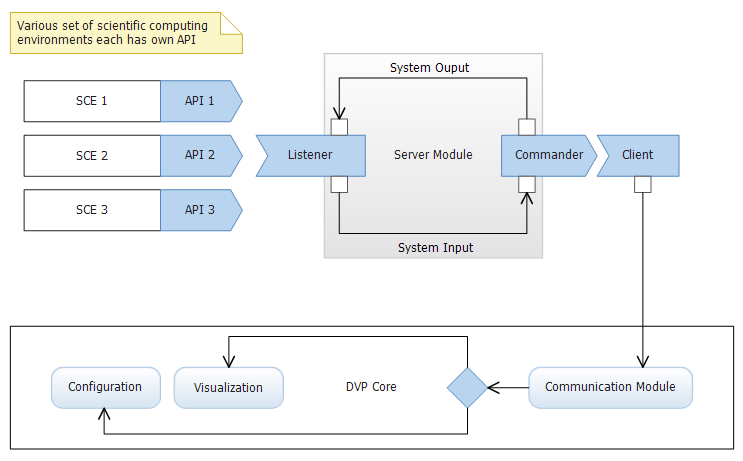}
  \caption{Communication server architecture}\label{FigCommManagerAct}
\end{figure}
A server module, as sketched in Figure \ref{FigCommManagerAct}, acts as a middle agent between DVP
and SCEs. This server module uses web sockets \citep{websocket} and XML HTTP
requests to establish communication between DVP and SCEs, by providing an interface for
communication with system. Since the server module is using the common web sockets, any SCE can be
seamlessly integrated with DVP by implementing a module that uses DVP interface provided by the
system module.

The server acts as a bridge providing an interface for the DVP visualization and configuration
functionalities to any other software. It even can communicate with regular programming languages
like C++, Java and C\#. Key SCEs, e.g., Matlab, Mathematica, and R, along with famous programmable
hardware chips, e.g., Arduino and Raspberry Pi, are planned to be supported for seamless integration
and communication with DVP.

%\subsubsection{System core} \label{sub:sysCore}
%\begin{figure}[t]\centering
%  \includegraphics[width=0.5\textwidth]{./Graphics/FigCoreArch.png}
%  \caption{System core architecture}\label{FigCoreArch}
%\end{figure}
%System core, as sketched in Figure \ref{FigCoreArch}, is the main logic layer where software core
%functions are implemented. It consists of a set of services and utilities that shape system work
%overflow. It also contains the implementation of different figures and visualization techniques. The
%main functions of this layer are to provide:
%\begin{itemize}
%  \item a set of data structures and algorithms that shape system behavior.
%  \item application communications with SCEs.
%  \item data sources services, including importing, modification and exporting.
%  \item validation and type checking routines for application data flows.
%  \item wrappers to UI for interfacing with figures, plots, and visualization techniques.
%\end{itemize}
%In addition, core functionalities include enormous set of sub routines used by UI application to
%utilize user interaction with application visual components.

\subsubsection{UI application} \label{sub:uiApp}
\begin{figure}[t]\centering
  \includegraphics[width=0.5\textwidth]{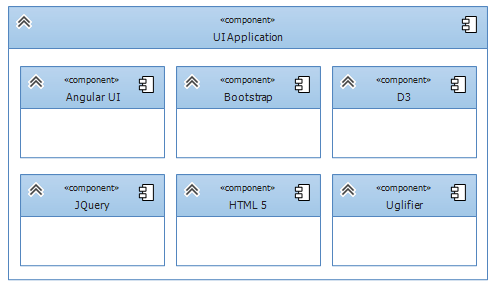}
  \caption{UI application components}\label{FigUIApp}
\end{figure}
UI application, as sketched in Figure \ref{FigUIApp}, is implemented as a web interface using a set
of web tools and a combination of the state-of-the-art work of leading companies and technologies,
e.g., Google, Twitter, and D3 \citep{d3js}. The later is a newly developed
library for dynamic and interactive web content. UI application will be hosted by the shell
application and they will act together as a single running application providing compatibility on
different operating systems and devices.

\subsubsection{Large Interactive Display Solution for Big Data}\label{SecBigDataDisplay}
\begin{figure}[t]\centering
  \includegraphics[width=0.5\textwidth]{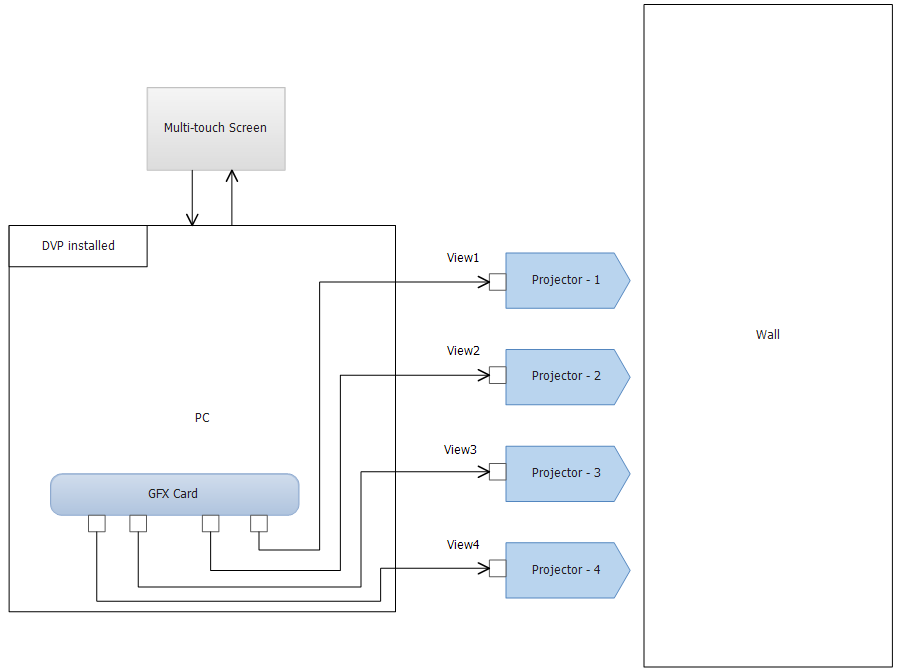}
  \caption{Architecture of integrated subsystem solution for big data
    visualization.}\label{FigBigDataDisplay}
\end{figure}
\begin{figure}[t]\centering
  \includegraphics[width=0.5\textwidth]{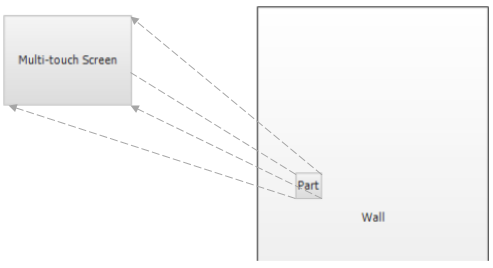}
  \caption{Big data is visualized on two displays simultaneously; one very large display for
    projecting the whole data, and another large touch screen for interacting with a portion of the
    data of particular interest.}\label{FigWall2Screen}
\end{figure}
A very large display subsystem to explore big data is to be designed. System architecture is shown
in Figure \ref{FigBigDataDisplay}. This system projects data on an area (a wall or any white screen)
covered by 4 parallel very high definition projectors. This area is estimated to be up to 100
m$^2$. Since data is huge, rendering is divided into 4 quadrants each is processed by a giant GPU
and fed into a separate projector. The whole rendered view is complied back by the 4 projectors to
the very large display screen.

Since, it is almost impossible to interact with data on that very large screen, another large touch
screen (80 inches) is connected to the system. If the analyst is interested in some portion of the
data displayed on the wall he can select it for interaction on the large touch screen, which can be
thought of as a magnifying glass in image processing softwares (Figure \ref{FigWall2Screen}).

\subsection{Features}\label{SecObjectives}
In this section, we provide and explain all the planned features that the architectural design of
DVP supports; we provide them in terms of user stories.
\subsubsection{Communication with Scientific Computing Environments (SCEs)}
DVP will provide a simple way to establish communication with scientific computing environments in
order to make a link between the process of data analysis and data visualization, and also to
facilitate reviewing the interpretation. A list of sub-features:
\begin{itemize}
  \item System can be attached and detached from SCE.
  \item Connection with single or multiple instances of SCEs.
  \item Connection can be made to SCEs within the same machine, over LAN, or even a remote machine.
\end{itemize}

\subsubsection{Manipulating data}
System has a set of features related to data manipulation that can be listed as follows:
\begin{itemize}
  \item Data imported will be categorized into one of three types: quantitative, categorical, or
  ordered categorical; this early categorization will help providing suggestion about the set of
  figures and plots that visualize data the best.
  \item DVP will support most common data format used by data analysts like CSV, XML, JSON or SGML.
  \item Data can be grabbed from online sources like Google drive, social media analytics, stock market, or even  other storage types, e.g., SQL or excel sheets.
  \item System will facilitate processing acquired using basic set of operations like merging datasets from various origins.
\end{itemize}

\subsubsection{Figures}
DVP provides a various set of figures and innovative visualization methodologies which have been a
result of research in the field of data visualization. A list of sub-features:
\begin{itemize}
  \item System will provide UI wizard to fill figure required parameters.
  \item Figures will have a UI selection way to create, modify, and delete groups of figure objects.
  \item Selection groups can be created using a mapping function to the data source.
  \item Annotations can be made to the figure itself or to one of its objects.
  \item Annotation will include creating arrows, circles, and polygons.
  \item Figure view can be transmitted to SCE.
\end{itemize}

\subsubsection{Extensibility}
Making DVP open for extensions will provide users with the power of implementing their own extension
that can help them with the process of visualizing or interacting with data, and build wider
community that uses DVP. This is achieved by providing a scripting language to facilitate a set of
key features:
\begin{itemize}
  \item Users can develop custom plots and it can be embedded to DVP.
  \item Also users will be able to implement transformation function that transform data before visualization.
  \item Users will be able to develop their own parameterized visualization methods with custom interactive behavior.
  \item Users can also implement post processors that can modify the result of visualization.
  \item Also all customizable visualization methods will be able to take custom pre and post callbacks as parameters.
  \item Figures will be reproducible even from the SCE by calling the written script.
\end{itemize}

\subsubsection{Interactivity}
Visual interaction with plots and models is a key part of visualization techniques. DVP provides a set of animated and interactive visualization methods, and most importantly provides to users a scripting language, as mentioned above, to produce their own interactive behavior. This feature is divided into: 
\begin{itemize}
  \item UI scroll bar can be used to control thresholds in different visual methods that make use of them.
  \item User can animate data points up to 2 million points in a non-interactive movie.
  \item User can use automatic or manual calibration to change constrains of data set sizes considered big or small.
\end{itemize}

\subsubsection{User interface}
DVP features related to UX can be listed as follows:
\begin{itemize}
  \item DVP will be cross platform, running on the three known operating systems Windows, Linux, and Mac.
  \item DVP will be able to run on most of tablets and smart phones.
  \item DVP can run on multi-touch screens.
  \item DVP will support running on multiple screens with providing a way to manage distributing figure among various screens.
  \item DVP will provide the ability to toggle between free and docked figure.
  \item DVP will be able to use pre defined set of hand gestures as a means of input.
\end{itemize}

\subsubsection{Exporting visual models}
DVP will provide methods to export visual models into the following forms:
\begin{itemize}
  \item Images with raster format supported by imagemagick.
  \item Images with SVG vector format.
  \item Interactive PDF document to encapsulate visualization with reporting.
  \item Print a single figure or multiple figure to any paper size.
\end{itemize}

\subsubsection{Web access}
The system will access web to do the following tasks:
\begin{itemize}
  \item Store user session information so that user can reload this session later from any machine.
  \item Save and share current visual models with colleagues or partners using web as if it is a web post.
\end{itemize}

\subsubsection{Business dashboards}
\begin{itemize}
  \item DVP will provide creating both local-machine and online business dashboards with multiple
  interactive linked figures.
  \item Dashboards created by DVP can obtain data from web or local data sources.
  \item Importing data from SQL storage servers to dashboards can be done with a UI wizard without writing any SQL code for the convenience of non-technical oriented users.
\end{itemize}

%%% Local Variables:
%%% mode: latex
%%% TeX-master: "Yousef2016DVP003"
%%% End:

\section{Global System Architecture and Design}\label{sec:architecture-design}
The main target of the design is to create an extensible, easy to use, beautiful visualization
platform that integrates seamlessly with all Scientific Computing Environments (SCEs). The current
DVP design consists of a Plugin written in Java for the intended SCE, a web server that uses JSON
serialization, and the DVP itself built with \emph{Web technologies} running on CEF (Figure
\ref{fig:currentDesign}). In this section, we first introduce the technologies used for the rest of
the paper (Section~\ref{sec:dvp:-design}). Then we go over several design alternatives and
approaches to document some design decisions that we have taken to finally reach this design of DVP.
\begin{figure}[t]\centering
  \includegraphics[width=0.5\textwidth]{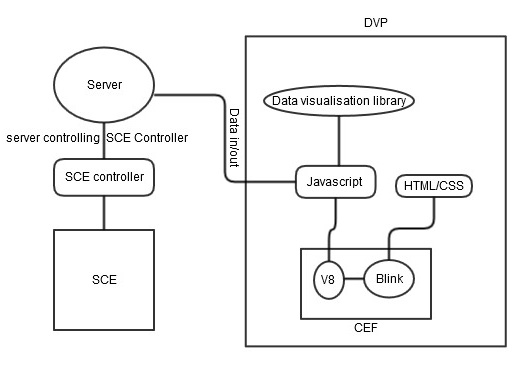}
  \caption{DVP architecture and design}\label{fig:currentDesign}
\end{figure}

\subsection{Technical background to DVP low level design}\label{sec:dvp:-design}
\subsubsection{D3js}
D3 \citep{Bostock2011D3DataDrivenDocument} is a Javascript library created by Mike Bostock
\citep{bostock} for visualization. It implements a Data-Driven Document model which works by
associating data with DOM elements, in order to facilitate common visualization tasks such as
selection, manipulation, addition and deletion of data points. It implements the model using a
syntax similar to that of jQuery \citep{jquery} which facilitates chaining several actions on a data
set in one line call. This ``chaining'' is not just syntax sugar, in fact it has a large effect on
performance because browsers can optimize consecutive rendering/relayout/repaint/restyle calls
\citep{restyle}.

D3 uses SVG standards introduced with HTML5 for drawing elements. SVG primitives are represented as
DOM elements, which allows for using any existing Javascript/CSS libraries to manipulate/style the
visualization elements. As an example, attaching an image to a data point represented as a circle in
a scatter plot would be as easy as just adding an \texttt{\textless img\textgreater} tag to the
circle DOM. Modern browsers can be seen as powerful and efficient rendering engines which is a fact
that D3 leverages. However, creating so many DOM elements causes problems with visualizations
requiring more than 200,000 elements as seen on our hardware and benchmarks; however, rendering
performance does scale with hardware.

\subsubsection{Google Chrome}
Google Chrome is a modern web browser built by Google. It is multi-threaded and has powerful
Javascript and rendering engines which leverage hardware acceleration through GPUs. Google Chrome is
based on an open source project called Chromium \citep{chromium}. There exists a sub-project called
Chromium Embedded Framework \citep{CFW} which is a framework for embedding Chromium-based browsers
in other applications. CEF is available as Dynamic Link libraries which one can use to build
applications. A new version of CEF is automatically generated for every new version of Chromium
through automated code extraction scripts. DVP is built using CEF.

Chrome, Chromium, CEF, all share the same codebase, and the same Javascript/rendering engines, they
are in fact at the core, the same thing. They only differ in GUI; Chrome comes with Gmail account
synchronization and some other Google services. Chromium has very similar UI but with some few
features stripped out. CEF is a DLL. The GPU accelerated compositing \citep{gpuacc} of Chromium and
its powerful rendering engine, Blink \citep{blink}, are the reason we currently use CEF.

\subsection{General Design Overview}
We believe that a client-server model should suffice for the needs of our system. The SCE acting as
the server, and a separate process called DVP which will run the visualization code acting as the
client. Both processes would then communicate through a serialization interface or Remote Procedure
Call (RPC). Following are some aspects to be considered:
\begin{enumerate}
  \item Due to the different nature of each SCE in terms of available datastructures, language and
  features, a special component or plugin should be built for each one but have a unified interface
  with a uniform serialization specifications to reduce the SCE specific code required in the DVP
  component. This component should be built with maximum re-usability in mind to avoid redundant
  design and coding.
  \item The DVP as a platform should be built with extensibility and re-usability in mind to account for the different scenarios of communication with different SCEs and user-added features/figures.
\end{enumerate}

\subsection{Comparison of Platforms}
\subsubsection{SCE Plugins}
SCE plugins can be written in different languages depending on each SCE design. But for the 3 main SCEs we target, namely, Matlab, Mathematica and R, it can be written in:
\begin{enumerate}
  \item The SCE Language itself.
  \item C++.
  \item Java.
\end{enumerate}
Writing the plugin in the SCE language itself means having to rewrite it from scratch for every SCE;
therefore having a high development and a high maintenance cost, and in some cases, inferior
performance to other 2 options. C++ is of course minimalistic and fast, however being native adds a
high maintainability cost, which can be avoided with Java, without sacrificing a lot of performance.
Depending on the DVP language and architecture, different serialization or RPC methods can be used,
each one has its pros and cons, however their impact thus far is not large. We discuss them and
their effects on our current design in Section \ref{DataSerialization}.

\subsubsection{DVP}
To build the DVP itself, we had to choose between different languages and platforms. It can basically be
built with anything, Java, C++, Javascript or even one of the SCEs. What follows is a discussion of
each platform and what it has to offer.
\paragraph{Java} has a cross-platform GUI framework, however its performance in visualization is
weak. Therefore to build a visualization tool using Java would require the use of OpenGL, or any
Java library or framework building upon it. There exists a programming language with a small
platform built for it for visualization using Java called Processing \citep{ulgdf}. It was developed
by MIT media labs. Even though it has a lot to offer, it was found that building unique interfaces
with rich features would require a lot of development effort using it, as compared to other
platforms like D3. Also, there exists a Javascript port for the language.
\paragraph{C++} is needless to say it offers the best performance when it comes to speed and
efficiency. However, both the development effort required to build the visualization library, the
rich interfaces mentioned in the user-stories, and in order to offer the level of extensibility
required, it would be very expensive and difficult to both design and build. However C++ may be
considered for visualizing large data samples or integrate with native components or libraries
needed by our system later on.
\paragraph{SCE Language} One could consider building the visualization platform using one of the
free SCEs, like R for instance and taking advantage of the familiarity of the users with it, and
take advantage of the primitives already existing. However, this would limit the scope of the
product to only scientific applications. Also this would limit the user-base to only those who know
that specific SCE and limit the performance of the visualization algorithms to that of the SCE's GUI
rendering engine.
\paragraph{Web technologies} The introduction of SVG standards in HTML5 and the evolution of
libraries such as D3, has given Javascript a large set of capabilities when it comes to
visualization. Modern browsers are very powerful rendering engines which enables Javascript to
render large datasets with ease and take advantage of hardware acceleration. Javascript is also
arguably the most popular programming language, which means a very large user base; and since
Javascript is an interpreted weak typed language, extensibility design would be very easy. The only
shortcoming of this approach is the access to native libraries such as OpenCV, or those needed to
integrate with certain hardware like Arduino for instance. However, Chromium and its embedded
framework provide a very feasible solution for this issue by allowing for Javascript to execute C++
functions through V8 and retrieve data in a native Javascript format. This means that Javascript can
access any memory available to C++. Another problem is handling massive data. By design, Javascript
arrays can only have 32 bit indices; this means it cannot handle data with more than 2 billion
points. But for such data, a native visualization library would be needed anyway for special
visualization. Therefore, a solution would be to have 2 modes: a full interactive mode and a simple
heavy visualization mode, where the heavy visualizations would be done through a separate native
libraries with specially optimized algorithms.

\subsubsection{Javascript and render engines}\label{comp}
If Web technologies will be used as a platform for DVP, this needs a further comparison among
Javascript/render engines. This comparison revealed that webkit is by far a head in performance if
compared to gejko.

\subsubsection{Platforms using Webkit and V8}
According to what is mentioned in Section \ref{comp}, we choose Webkit and v8 to be the DVP
engines. This give us another set of options to be compared below.
\paragraph{Google Chrome}
\begin{description}
  \item[pros] \hfill
  \begin{itemize}
    \item no build problems.
    \item more portable.
  \end{itemize}
  \item[cons] \hfill
  \begin{itemize}
    \item no shared memory.
    \item no advanced integration with javascript since it is closed box run on the web browser.
    \item cannot handle big data because of lake of control over it.
    \item no security.
    \item restrictions on javascript access.
  \end{itemize}
\end{description}
\paragraph{Chromium}
\begin{description}
  \item[pros] \hfill
  \begin{itemize}
    \item every con in Google Chrome, above, can be alleviated here.
  \end{itemize}
  \item[cons] \hfill
  \begin{itemize}
    \item maintenance problem with the update as we must remove google things every time to be up to dated.
    \item cross platform build requirements.
  \end{itemize}
\end{description}
\paragraph{CEF}
\begin{description}
  \item[pros] \hfill
  \begin{itemize}
    \item every con for Google Chrome, above, can be alleviated here.
    \item no maintenance problem.
  \end{itemize}
  \item[cons] \hfill
  \begin{itemize}
    \item cross platform build requirements.
  \end{itemize}
\end{description}
\paragraph{CEF variants} Many development platforms have been built around it, some for desktop
development, some for server development, and even some for mobile development. We have decided to
go with native simple CEF; however below we keep the discussion and list other forks that we
considered.
\paragraph{Crosswalk \citep{crosswalk}} is maintained by Intel. It is mainly intended for mobile
development and that is where their main development and support goes; therefore it is a viable
option when developing the mobile interface.
\paragraph{Awesomium \citep{awesomium} \& Tide SDK \citep{tidesdk}} Awesomium is a commercial
library, built on top of CEF; it offers an SDK with some built-in functionality. It is just like
TideSDK; however building on top of such libraries exposes us to bloat, less flexibility in design,
and restrictions. E.g., we will be neither able to update nor develop using new features of CEF as
soon as they come out (see Section \ref{encodingProblems} for a real usecase).
\paragraph{NodeJs \citep{nodejs}} is a headless V8 engine with no rendering engine such as
webkit. Of course this cannot be used directly for visualization; however NodeJs, can act as a very
strong server with rich access to native libraries since NodeJs already integrates with a lot of
native libraries. However, integrating it with CEF is very difficult and done by applying patches to
direct9 source code such as cefode project \citep{cefode}.
\paragraph{Node Webkit \citep{nodewebkit}} integrates NodeJs with Chromium. Unfortunately, Node
Webkit has its own ``packaging'' system and your app is ``packaged'' into a special zip format, and
it has its own GUI library that builds on top of that of Chromium. Modifying it would be a lot of
work. Using most of the libraries munitioned above would offer little help with the DVP development;
however the cost of maintainability and control over CEF components and updating would be too
high. That is why we resorted to using plain CEF since it meets all the DVP design philosophy
aspects.

%%%%%%%%%%%%%%%%%%%%%%%%%%%%%%%%%%%%%%%%%%%%%%%%%% 
%%%%%%%%%%%%%%%%%%%%%%%%%%%%%%%%%%%%%%%%%%%%%%%%%% 

\section{Chromium Embedded Framework (CEF)}
This section discusses CEF and all its related issues in depth. This section provides many technical
details that seems to be out of scope of the present article. However, we would like to illustrate
them because they very much relate to the high level objectives and philosophy of DVP\@. So, down
the rabbit hole we go.

\subsection{Getting Started}
The CEF project is overall poorly documented; however, project wiki \citep{wiki} is a good place to
start. Some tutorials, e.g., \citep{tutorial1,tutorial2}, provide a good introduction to most of the
functionalities provided by CEF\@. Some forums, e.g., \citep{forum}, is also a good resource for
help. However, for too many hits questions are left unanswered.

Upon downloading and extracting the binary packages, the sample application can be built using the
provided \texttt{build.sh} file on Linux or using the \texttt{sln} file on Windows. These project
files are automatically generated through ``Generate Your Project'' (GYP) \citep{GYP} scripts;
editing these files is quite tedious. Also, editing the GYP files is not possible without
downloading almost all of Chromium source. For those reasons, we decided to build our own build
scripts using \texttt{CMake} \citep{cmake} to allow for consistent cross-platform build.

\subsection{Building Upon CEF}
CEF comes with two sample applications: CefSimple and CefClient. The first is a minimalistic window
with a single frame. The second implements almost all of CEF features. We decided to build upon
the CefSimple instead of stripping out the CefClient.

Both of CEF sample applications have 3 versions; one for each of Windows, Mac and Linux. It uses
native window toolkits on each of those platforms. We wanted to build a cross-platform application;
and doing so requires building the GUI 3 times. We know that, by DVP design, the UI will mostly rely
on Javascript; however, in the future when adding heavier methods that require, e.g., OpenGL, we
might need to build some native controls outside the CEF frame.

Therefore, we decided to use wxWidgets, which is a library built as an abstraction layer on top of
native libraries of each platform: windows forms for Windows, GTK for Linux, and Cacoa for
Mac. However embedding CEF still requires a native applet, e.g., gtk\_vbox, which required
digging inside wxWidgets and finding a control that exposes such an applet and extracts a reference
to it. This is discussed further in Section \ref{wxWidgets}.

Also, for some unknown forsaken reason, CEF includes its header files using relative paths,
i.e. quotation. This destroys any dreams of building outside the directory of CEF since it requires
maintaining the hierarchy where the includes are located. As a solution, we created a CMake file in
intended directory, and built everything relative to that path. However, CMake supports out of
source directory build, so everything is built into \_\_\texttt{build}.

\subsection{Encoding Problems}\label{encodingProblems}
After CEF is embedded into wxWidgets, we wanted to integrate it with other libraries such as OpenCV
to be accessible from Javascript. This required delving into a larger problem, which is
encodings. Javascript is not designed to handle binary data. Although typedarrays \citep{typedarray}
do exist in Javascript and are supported by all of Chrome variants including CEF, which does not
expose an interface for creating these arrays from C++ \citep{ceforum}. The best way to pass binary
data, as advised by Google support, is surprisingly by creating an XHTML request, and catch it
\citep{ceforum}. The available options for passing binary data are through:
\begin{description}
  \item [an XHTML request,] which is the worst option in terms of performance, design and
  implementation.
  \item [encoding into strings,] which is implied by the first option. It is similarly bad in
  performance but not in design.
  \item [converting everything into normal float arrays,] then passing that instead. This is
  feasible, with acceptable performance. However, the overhead of copying data can sometimes be
  unacceptable if the data size is large.
  \item [direct Proxies \citep{proxies},] which is a sort of overloading the Javascript
  \mbox{\texttt{[]}} operators using a C++ function; i.e. when \mbox{\texttt{arr[0]}} is executed, a
  C++ function is called with the parameter ``index'' with value 0, so we can really make Javascript
  access any block of memory available to C++ using pointers. Now, Javascript arrays are C++
  pointers. That is really cool, but not so fast!
\end{description}

Proxies are a part of ECMA 6, and that is not fully implemented nor supported in Chrome variants at
the time of releasing the first version of DVP \cite{v8issues}. However, incomplete support for it
is provided through an experimental feature, which is turned on by passing the
\mbox{\texttt{--harmony}} flag to V8 and using a library \citep{v8issues2} to account for the
missing features. However, this should be fixed in later versions after the launch of ECMA 6.

\subsection{Integrating Libraries with CEF}
\subsubsection{wxWidgets}\label{wxWidgets}
As mentioned above, wxWidgets is an abstraction layer built on top of each platform native window
toolkit. This means that using wxWidgets requires using the native windows at the end of the
day. However, if, at some point, a direct access is required into the native toolkit objects, a hack
or extend is needed. This hacking/extending will be a platform specific code. Since hacking solution
was not only disastrous but also quite tedious (since almost none of wxWidgets panels uses
\texttt{gtk\_vbox} on linux), we decided to extend it. However, even that required a little hacking
of its own since its extension is neither documented nor standard.
\subsubsection{wxWidgets render Loop}
CEF requires a blocking call for its event and render loops to start; this call blocks until CEF is
shutdown. This caused a problem with wxWidgets because it required the need for launching another
thread for that function to be called from. The optimal solution however is to integrate CEF loop
with wxWidget render loop.  wxWidgets starts its render loop using a macro; however, it can be
overridden to let your own render loop see Making render loop \citep{loop}. This required a lot of
work. Rather, we proceeded by starting the CEF thread in an event call. Events are multi-threaded
but safe(r) when interacting with wxWidget GUI components \citep{wxwidgets1}; until now, no
complications have occurred.
\paragraph{Extending wxWidgets on Linux} After diving into the abyss of wxWidgets implementation for
linux, we emerged with the fact that there exists a variable named \texttt{m\_widget}, which
contained the actual GTKWidget being used by any control that inherits from wxControl. However, some
solution is already available and posted on wxWidgets forums \citep{wxWidgets}.
\paragraph{Building Problems on Linux} wxWidgets is built using AutoTools, then installed into a
folder of choice with symbolic links to compiled files. We wanted to specify the build to use the
wxWidgets build available in our repository instead of that coming with linux distros. There exists
a CMake variable called \texttt{wxWidgets\_ROOT\_DIR}, which turned out to work only on Windows. In
addition, it turns out that wxWidgets comes with its own \texttt{wx-config} file which is similar to
\texttt{pkg-config}. Therefore, it is necessary to specify this file using the CMake variable
\texttt{wxWidgets\_CONFIG\_EXECUTABLE}. Another issue is that wxWidgets adds an \texttt{isystem}
flag to the compiler parameters automatically for some reason. This of course destroys any attempt
to compile. The only way to disable it is setting the variable
\texttt{wxWidgets\_INCLUDE\_DIRS\_NO\_SYSTEM} to true.

\subsubsection{OpenCV}
There are two ways to export OpenCV into Javascript: (1) by making a full fledged interface, with
matrix objects that resembles OpenCV \texttt{Mat} class (this is the way nodeJs OpenCV package does
it), (2) or by making a limited set of functionality that is called through functions. By the time
of releasing the first version of DVP, we have only created a limited set of functions as a proof of
concept. Encoding images is also one of the main issues with OpenCV, since Javascript can only
handle images encoded in \texttt{base64} strings, which required using external libraries to encode
such images. However, an alternative solution \citep{canvassolution} is to use a canvas to display
images, and supply it with a typedarray. Also, this can possibly be mixed with a Proxy to avoid
copying, which we never investigated yet.

%%%%%%%%%%%%%%%%%%%%%%%%%%%%%%%%%%%%%%%%%%%%%%%%%% 
%%%%%%%%%%%%%%%%%%%%%%%%%%%%%%%%%%%%%%%%%%%%%%%%%% 

\section{Javascript Application}
DVP is a standalone application that we had planned to write using web technologies (HTML5,
CSS/CSS3, and Javascript) and C++ as its backend using CEF integration. As we mentioned earlier, our
general architecture for DVP is a client-server model. This should not lead to the wrong conclusion
that DVP cannot run without server side component. It should be very clear that client-server
architecture has been chosen to handle SCE integration with DVP; in addition, DVP can run as
standalone application without the need for any server side component. On the other hand, had we
opted to use something like nodeJS as our backend, DVP would become client-server application. In
this section we will compare the different Javascript frameworks and different ways for figure
drawing; also, we will elaborate on architecture and on how user requirements are fulfilled.

\subsection{Comparison of Web Technologies}
\subsubsection{Javascript, AngularJS, and HTML}
HTML is great for declaring static documents; however, it falters upon using for declaring dynamic
views in web-applications. AngularJS allows for extending HTML vocabulary for applications. The
resulting environment is extraordinarily expressive and readable with quick develop rate. Other
frameworks deal with HTML shortcomings by either abstracting away HTML, CSS, and/or JavaScript on a
hand; or by providing an imperative way for manipulating the DOM on the other hand. Neither of these
addresses the root problem of that HTML was not designed for dynamic views. AngularJS is a toolset
for building the framework that is most suited to application development. It is fully extensible
and works well with other libraries. Every feature can be modified or replaced to suit a unique
development workflow and feature needs. For more details check angular website \citep{angular}

\subsubsection{HTML5 Draw types and libraries}
\paragraph{Scalable Vector Graphics (SVG)} is used to define vector-based graphics for the Web
without sacrificing any quality if they are zoomed or resized. Every element and every attribute in
SVG files can be animated; it is easy to make very high interact-able figure.
\paragraph{Canvas} is a raster graphics that is mush faster in drawing; however, it will be too hard
to make high interactive figures with it.
\paragraph{When SVG? When Canvas?} We decided that we will make two flows for our application: one
using SVG that will be very highly interactive with full features but will have limitations on data
size. The other flow will use canvas and will have limited interactive figures but can handle huge
data size.
\paragraph{Why D3?} for the SVG flow, D3 is a mature library for drawing interactive figure with
very rich ready-made examples and figures.

\subsection{Architecture}
The application-side architecture should not be confused of course with the architecture discussed
in Section \ref{sec:architecture-design}. There are three main components of the application-side
architecture:
\begin{description}
  \item[User Interface (UI)] should provide excellent user experience. Its implementation relies
  heavily on mastering:
  \begin{itemize}
    \item Jquery, CSS, HTML.
    \item Angular and data binding.
  \end{itemize}
  \item[Core] is related to how we mange the underlying layers of application like validation, SCE
  integration, CEF integration, etc. Its implementation relies heavily on mastering:
  \begin{itemize}
    \item javascript.
    \item Object Oriented Design (OOD).
    \item dependency injection design pattern.
    \item Angular and its services.
  \end{itemize}
  \item[Figures] is related to how the figures themselves, which the user will interact with, are
  created. Their implementation relies heavily on mastering:
  \begin{itemize}
    \item D3.
    \item Jquery.
    \item Angular
    \item how render engines works.
    \item data visualization foundations, e.g., geometry, probability statistics, linear algebra,
    etc.
  \end{itemize}
\end{description}

\subsubsection{Core}
This component should:
\begin{itemize}
  \item manage the underlying layers of the application that the UI reflects.
  \item keep the system in consistent state by validating every input/output action and by providing
  both the data structure and logic to handle this.
  \item integrate with anything outside the javascript application, e.g., SCE or CEF.
\end{itemize}
Therefore, and because of the lack of classes and OOP in javascript, we had to take care of several
technicalities as follows.
\paragraph{Data Structure} Simple lists have javascript objects; each object has a definition for
its fields and a unique filed that is used to store/access/delete the object in the list. The unique
field name will be given as a parameter in the list constructor. In principle, although not needed
so far, the object may have a field that needs a special data structure.
\paragraph{Validation} we need to construct a lot of things:
\begin{itemize}
  \item object duplication.
  \item definition for each object type; e.g., data-source object definition is
  \texttt{id, name, data, ColumnNames, ColumnTypes}.
  \item validation that each object related to specific type has the required fields to be added in
  the system.
  \item validation of required fields values.
\end{itemize}
\paragraph{OOP} we need to reinvent almost everything in this regards. \texttt{Validation} class has
all the required functions to validate an object according to a given set of
parameters. \texttt{List} class, a service in angular, can add/delete/modify/validate the Object to
the list. This is in addition to some other utilities, e.g., checking for element existence. Its
constructor takes object definition, key field name, and list name as parameters. A class X will
inherit the class List then add/overload other functions if needed %the List class functions to do the specific tasks related to what x should handle, then call the super function to do the rest of usual work like validating the Object with its definition given in the constructor of the x class. x class can be Datasource, Figures, SCEs, etc.
\paragraph{SCE integration} For integration with SCEs, we simply need to establish the two-way
communication between DVP and the SCE. Therefore, we need to send commands from DVP (or SCE) to SCE
(or DVP). As mentioned earlier we are following client/server model to handle this communication
using JSON objects passed between DVP and SCE. From DVP side we use AJAX functions from JQuery to
handle post requests and SSE to handle get requests. Alternatively, the angular service, HTTP, could
be used; however, we found that the SCE plugin does not work with this approach, which may need more
debugging.

\begin{description}
  \item[To receive commands] from core architecture mentioned above, e.g., to create a figure, it is
  only required to get information from the user, construct it in a javascript object, and pass it
  to the core to be validated and added. Therefore, to receive commands from SCE, there will be a
  protocol to define the kind of operation needed then pass the object constructed, by the SCE
  plugin (Section~\ref{chp:SCE-Pluging}), from the user input to the related function in the core;
  e.g., \mbox{\texttt{figure.add(constructedObject)}}. The protocol is also an object definition
  since it is a JSON object.

  % \begin{itemize}
  %   \item Operation.
  % \end{itemize}

  \item[To send commands] this is established via ajax post function, which is very
  straightforward.%nothing to mention here all is work is related to SCE pluging.
  % \paragraph{CEF integration} We did not finish it and omar should told sameh about it but this is a
  % keywords:
  % \begin{itemize}
  %   \item Proxies (for data sharing).
  % \end{itemize}

\end{description}
%%%%%%%%%%%%%%%%%%%%%%%%%%%%%%%%%%%%%%%%%%%%%%%%%% 
%%%%%%%%%%%%%%%%%%%%%%%%%%%%%%%%%%%%%%%%%%%%%%%%%% 

\section{SCE Plugin}\label{chp:SCE-Pluging}
\subsection{Data Serialization Options}\label{DataSerialization}
Since Javascript cannot handle binary data, some sort of serialization is necessary. However,
serialization could deteriorate performance with larger datasets. Several serialization methods are
possible:
\begin{itemize}
  \item JSON and its binary variant BSON.
  \item Google Protocol Buffers \citep{protobuf}.
  \item Protocol Buffer's Author's Cap'n Proto \citep{capnproto}.
  \item Previously Facebook, now Apache, Thrift \citep{thriftappache}.
  \item Apache Aciteve MQ \citep{apacheacitive}.
  \item Shared Memory.
\end{itemize}
Each option has its own pros and cons. The first option, JSON, is the one we currently use. It is
the most straightforward and native to Javascript. In addition, we do not observe performance issues
at the moment; the time consumption is still acceptable at a matrix size of 200,000$\times
$50. Other libraries use specific formats; porting them to handle Matlab objects is
cumbersome. %However, there is a lot of ground for optimization by even mixing some of these formats.
On the other hand, Cap'n Proto seems very promising since it does not serialize; rather, it just
copies bytes in a cross-platform
manner. %Also, having CEF, with all the C++ libraries in the world available, pretty much any serialization format or even shared memory can be used. But again, this is fine tuning, and we have not spent much time on it at the moment since JSON is meets our requirements.

\subsection{Current Implementation}
\subsubsection{Overview}
We have used Jetty because it is minimalistic, small, extensible, and sufficient for our
purposes. The designed servlets architecture is meant to be as much extensible and reusable as
possible. Adapting these servlets for any SCE should be trivial as discussed in section
\ref{SCEDetails}. For Matlab control, there are many approaches \citep{wikiapproaches}. We have
decided to go with ``MatlabControl'' since it is simple, straightforward, and it works on existing
opened sessions without the need to open a new session.

\subsubsection{Some low-level details}\label{SCEDetails}
\paragraph{Component Structure} \texttt{DVTWebServerInterface} is the entry point of the component;
it can be extended to modify logic or add functionality for any SCE type if desired. E.g., we have
extended it using \texttt{MatlabDVTServerInterface} only to call the constructor with the
\texttt{\$sceName} parameter and to specify the server post service \texttt{sse} and to specify the
default \texttt{\$serializationType} (json) since java dose not support compiler
directives. \texttt{DVTWebServerInterface} initializes 4 servlets: \texttt{DVTNewDVTIdServlet},
\texttt{DVTSCEJsonServlet}, \texttt{DVTEventSourceServlet} and \texttt{DVTEventSourceReplyServlet};
then maps them to the urls \texttt{welcome}, \texttt{\$sceName}, \texttt{sse}, \texttt{sse-reply}.

\texttt{DVTNewDVTIdServlet} is responsible for handshaking with DVP and sending configurations to it
including the urls to which all other servlets are mapped. It sends an object of
\texttt{NewDVTIdMessage} from messaging serialized in \texttt{\$serializationType}. It is also
responsbile for assigning IDs to DVP; however, this is done automatically when creating a new
\texttt{NewDVTIdMessage}; the ID of the message is thread-safe auto-incrementing.

\texttt{DVTSCEJsonServlet} is responsible for all interfacing with SCE, connection, disconnection,
evaluation, and storing variables. It does not need to be extended, e.g., to add support for other
SCEs. Rather, simply implement \texttt{SCEJsonInterface} and inject it into the constructor. It
talks to DVP through an \texttt{SCEEvalMessage} from messaging serialized in JSON. One possibility
is to make an abstract parent class and extend it to save any redundant code in several SCE
communication servlets implementing different \texttt{\$serializationType}.

\texttt{DVTEventSourceServlet} is responsible for registering SSE connections with and sending SSE
events to DVP. It extends jetty implementation of SSE servlet (located at
\texttt{./org/eclipse/jetty/servlets/EventSource*.java}). When an SSE request is sent, it is
assigned a thread-safe ID and added to a thread-safe list. \texttt{DVTEventSourceReplyServlet} uses
this ID to check for replies to this specific request.

\texttt{DVTEventSourceReplyServlet} is responsible for accepting replies to SSE requests and doing
blocking waits for the replies (if required) using thread-safe lists and IDs supplied to it and
obtained from \texttt{DVTEventSourceServlet}. Yet, another important to-do task is to make sse
message an sse json message. Next, for more clarification, we provide a simple scenario for SSE
workflow.
\begin{enumerate}
  \item SSE initialization is requested by DVP; eclipse implementation calls the function
  \texttt{newEventSource}, which is an abstract and implemented in this class, and the DVP is added
  to \texttt{\$eventSources} list in \texttt{DVTEventSourceServlet}.
  \item SSE request is sent through the function \texttt{sendDataToDVTClient} in
  \texttt{DVTEventSourceServlet}, the request is assigned a static thread-safe ID that is created
  once an instance of \texttt{SSEMessage} from messaging is created.
  \item The SCE calls \texttt{waitForSSEReply} in \texttt{DVTEventSourceReplyServlet}, providing it
  with the request ID returned from \texttt{sendDataToDVTClient} and the \texttt{DvpId}. The
  servlets blocks with a timeout, checking for the thread-safe list for the request arrival.
  \item DVP sends a reply to the \texttt{sse-reply} url, and an instance of \texttt{SSEReplyMessage}
  is created and added to the thread-safe list, which is checked in the \texttt{waitForSSEReply}.
\end{enumerate}

\texttt{SCEJsonInterface}, an interface with SCE, provides connection, disconnection, evaluation and
storage functionalities. It is required for use in \texttt{DVTSCEJsonInterfaceServlet}. Another
to-do task is to make an abstract parent and extend it to handle other
\texttt{\$serializationType}. Classes/Interfaces needed to be implemented/extended to add
functionalities for other SCEs.

% \begin{enumerate}
%   \item \texttt{SCEJsonInterface} (e.g. \texttt{MatlabJSONInterface}, inject its implementation at
%   \texttt{DVTSCEJsonServlet}'s constructor).
%   \item \texttt{DVTWebServerInterface} (optional, e.g. \texttt{MatlabDVTServerInterface} this should
%   be ur entry point).
% \end{enumerate}

%%% Local Variables:
%%% mode: latex
%%% TeX-master: "Yousef2016DVP003"
%%% End:

\section{Conclusion and Future Work}\label{sec:concl-future-work}
This article presented the design and implementation of DVP, a Data Visualization Platform, with an
architecture that attempts to bridge the gap between the Scientific Computing Environments (SCE)
that analyze data and Data Visualization Software (DVSW) that visualize data. DVP is designed to
connect seamlessly to SCEs to bring the computational capabilities to the visualization methods in a
single coherent platform. DVP is designed with two interfaces, the desktop standalone version and
the online interface. A free demo for the online interface of DVP is available \citep{DVP}. Although
the architecture of DVP is flexible to allow for integration with any SCE, the current
implementation is only provided for Matlab. The future version of DVP is an open-source version that
integrates with Python to provide wider support for the whole Python community, in general, and for
the ``Data Science'' community in particular.

%%% Local Variables:
%%% mode: latex
%%% TeX-master: "Yousef2016DVP003"
%%% End:

\bibliographystyle{IEEEtran}
\bibliography{publications,booksIhave,URLs}

\clearpage

\appendices
\section{Background, Tutorials, and Motivation From Textbooks}\label{sec:backgr-tutor-motiv}
This section is a very short account and tutorial on data visualization and graphics, taken almost
from textbooks, that includes history, importance and motivation for exploratory data analysis and
Grammar of Graphics (GoG). It is important for a reader who is not fully acquainted with the field;
however, it can be trimmed out from the manuscript without affecting its coherence.
\begin{figure}[p]\centering
  \includegraphics[width=0.4\textwidth]{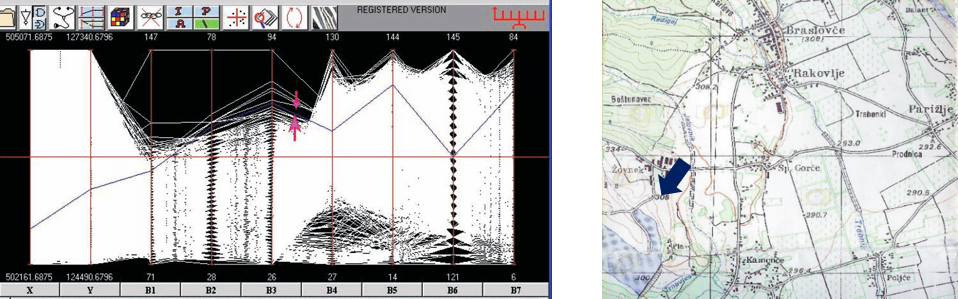}

  \bigskip

  \includegraphics[width=0.4\textwidth]{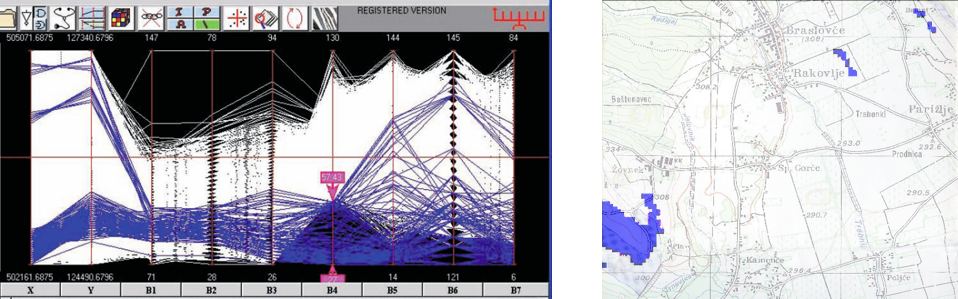}

  \bigskip

  \includegraphics[width=0.4\textwidth]{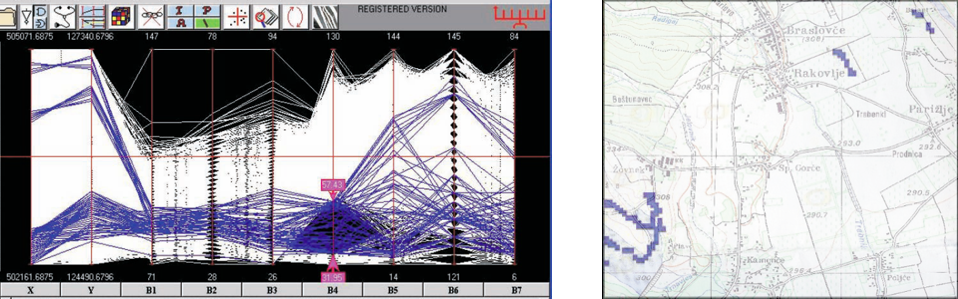}

  \caption{ Borrowed with little modification from \cite[Sec. 10.2.2]{Inselberg2008ParCord}. First
    row: map of Slovenia (right), and its satellite data plotted on ||-coords (left). The blue line
    selected on the ||-coords represents a point on the map with two coordinates and 7 sateliate
    readings. Second row: a set of observations with visual interest is selected on the ||-coords
    (left) surprisingly corresponds to the location of water on the map (right). Third row: a set of
    observations with visual interest is selected on the ||-coords (left) surprisingly corresponds
    to the location of the boundary of water (right).}\label{FigInselberg2008ParCordSolevania1}
\end{figure}

\begin{figure}[t]\centering
  \includegraphics[width=0.4\textwidth]{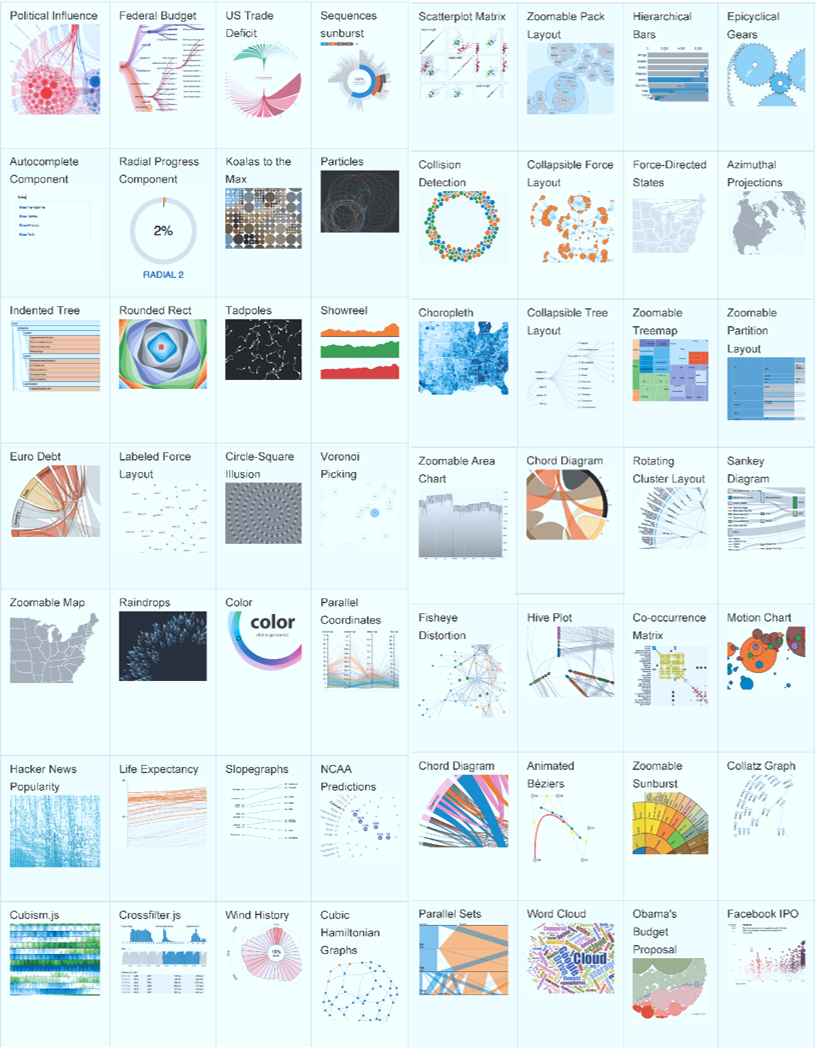}
  \caption{Some data visualization methods important to different fields of science. Interactive
    version for demonstration is available at \cite{D3gallery}}\label{FigManyPlots}
\end{figure}

\begin{figure}[t]\centering
  \includegraphics[width=.4\textwidth]{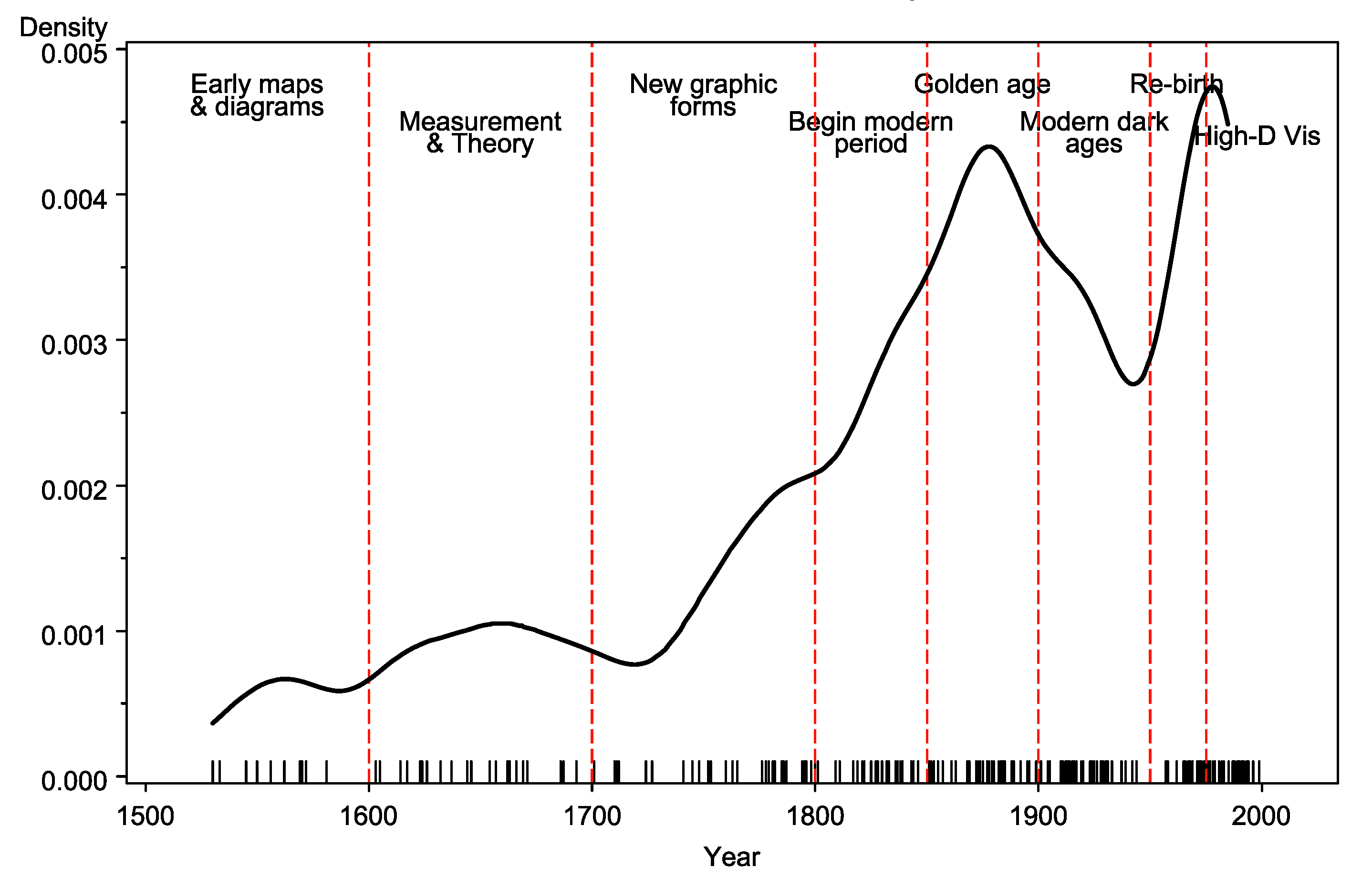}
  \caption{Borrowed, \citep[and appears in][as Figure 1.1]{Chen2008DataVisualization}. Time
    distribution of events considered milestones in the history of data visualization, shown by a
    rug plot and density estimate.}\label{FigChenMilestoneDensity}
\end{figure}

\begin{figure}[t]\centering
  \includegraphics[width=.4\textwidth]{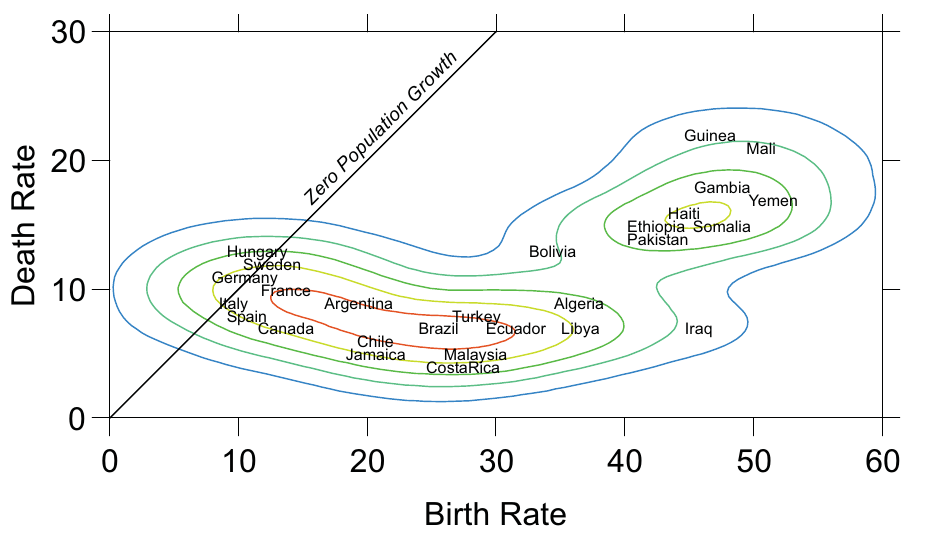}
  \caption{Borrowed from \cite{Wilkinson2006GrammarOfGraphics}. Plot of death rates against birth
    rates for selected countries.}\label{FigWilkinsonDeathBirth}
\end{figure}

\begin{figure}[t]\centering
  \includegraphics[width=.4\textwidth]{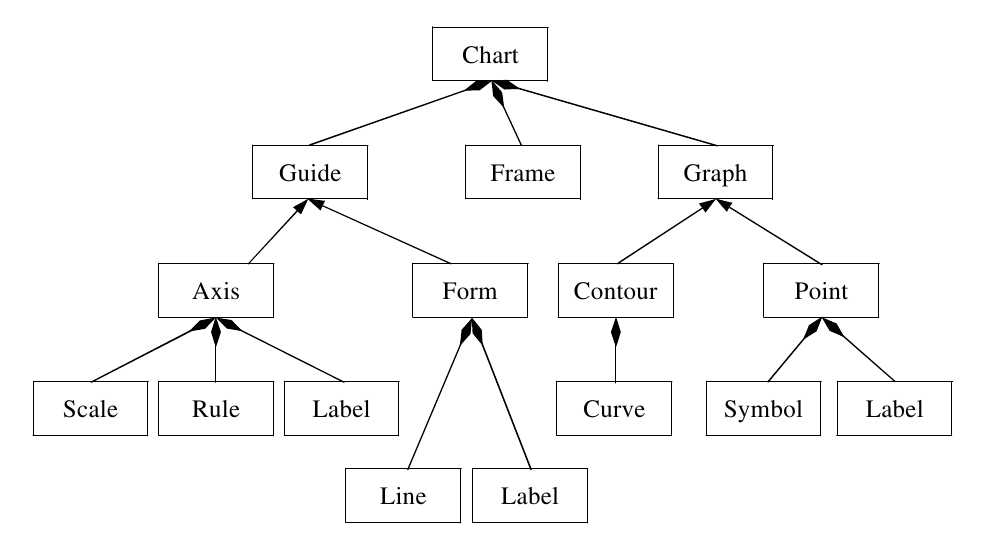}
  \caption{Borrowed from \cite{Wilkinson2006GrammarOfGraphics}. Design tree for chart in Figure
    \ref{FigWilkinsonDeathBirth}.}\label{FigwilkinsonDesignTree}
\end{figure}

\begin{figure}[t]\centering
  \includegraphics[width=.4\textwidth]{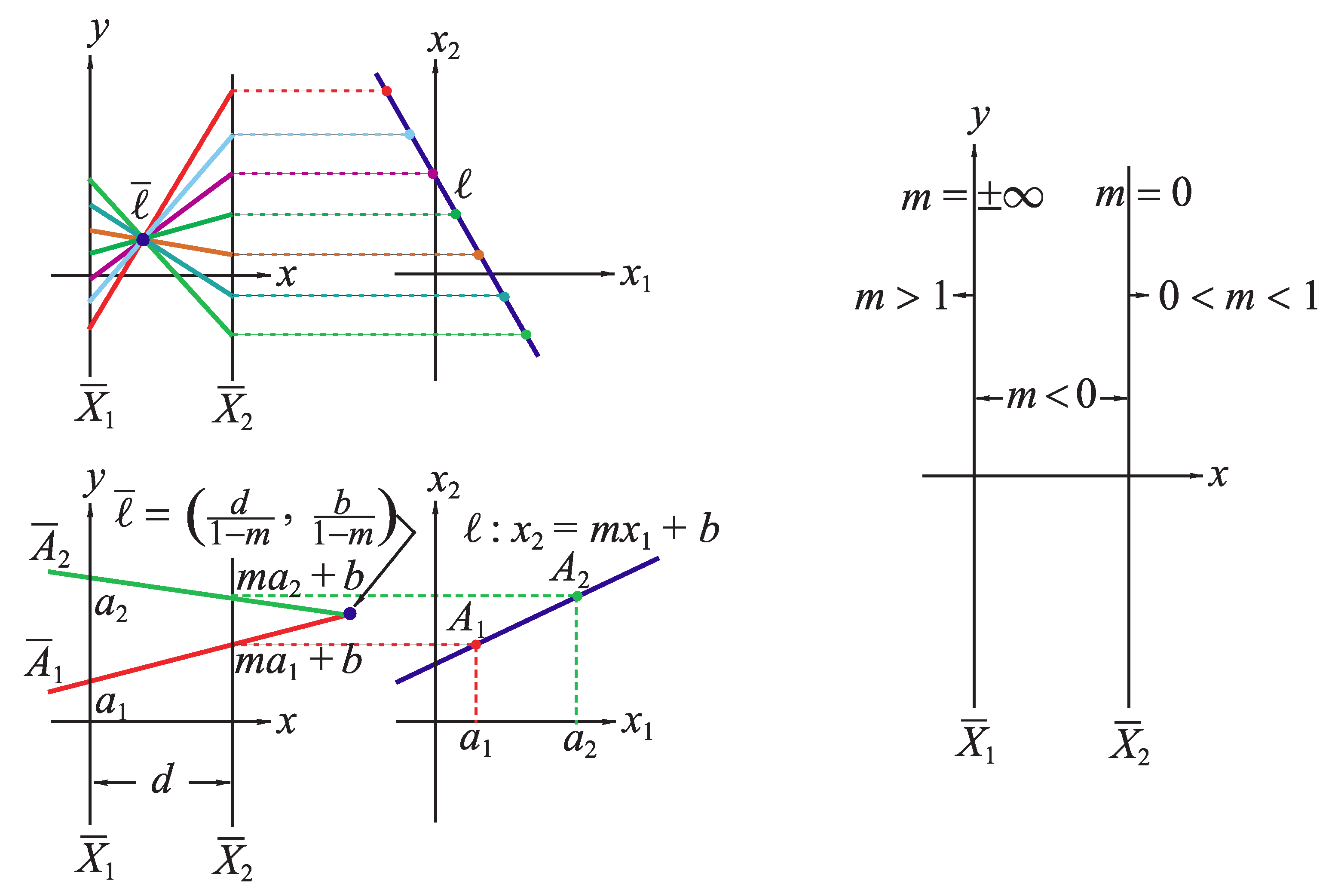}
  \caption{Borrowed, \citep[and appears in][Sec. 10.2.2]{Inselberg2008ParCord}. Geometry of 2D line
    in both perpendicular and parallel coordinates.}\label{FigInselbergGeomParCord}
\end{figure}

\subsection{History and Evolution of Graphics}\label{sec:history-graphics}
Figure \ref{FigChenMilestoneDensity}, \citep[which appears in][as Figure
1.1]{Chen2008DataVisualization} provides a graphic overview of the evolution of data visualization
presented as density of major developments in the field over time. The epoch of 1850--1900 was named
the ``golden age'' for the many innovations in graphics and thematic cartography that took place for
understanding data. The epoch of (1900--1950) was named ``modern dark age'' for the decline in
graphics and visualization development as a result of the rise of quantification and formal models
and tendency to quantize and formalize things. The epoch of 1950--1975 was named the ``rebirth of
data visualization'' as a result of the great developments known in the literature by Exploratory
Data Analysis (EDA) that connects visualization to analysis and quantification. The epoch of
1975--2000 was named ``high-D interactive and dynamic data visualization'' for invention of many new
methods of visualization, interaction, new methods of visualizing high dimensional data, etc.

On the other hand, ``\textit{Computing advances have benefited exploratory graphics far more...The
  importance of software availability and popularity in determining what analyses are carried out
  and how they are presented will be an interesting research topic for future historians of
  science...In the world of statistics itself, the packages SAS and SPSS were long dominant. In the
  last 15 years, first S and S-plus and now R have emerged as important competitors. None of these
  packages currently provide effective interactive tools for exploratory graphics, though they are
  all moving slowly in that direction as well as extending the range and flexibility of the
  presentation graphics they offer.}''\citep{Chen2008DataVisualization}. We add Matlab and
Mathematica, which are two very important and powerful data analytic software, to this list.

\subsection{Exploratory Data Analysis: importance and example on}\label{sec:expl-data-analys}
This section conveys both the scientific need and the financial opportunity for a sound and
elaborate data visualization software. We borrow Figure \ref{FigInselberg2008ParCordSolevania1} with
little modification from \cite[Sec. 10.2.2]{Inselberg2008ParCord}. This dataset is part of hyper
spectral satellite data for a portion of Slovenia, in Europe. The map of that portion is on the
right of Figure \ref{FigInselberg2008ParCordSolevania1}. The dataset consists of 9 dimensions and
9,000 observations. Each observation represents a point on the map with 2 dimensions (named
\textit{X} and \textit{Y}) for its location; the other 7 dimensions (named \textit{B1}--\textit{B7})
are data collected from satellite measures for that particular point.

The first aspect of good data visualization is the ability to view data in dimensions higher than
three! Figure \ref{FigInselberg2008ParCordSolevania1} (first row left) is a parallel-coordinate plot
\citep{Wegman1990HyperdimensionalParCord,Inselberg2008ParCord} for this dataset produced by
Parallax, the commercial software of the author of \cite{Inselberg2008ParCord}. In ||-coords, axes
are located parallel to each other as opposed to the perpendicular Cartesian coordinate system. A
point in ||-coords is represented as connected line segments, which intersect with variable axes at
the corresponding feature values of that point. For example, on Figure
\ref{FigInselberg2008ParCordSolevania1} (first row right) the point on the map pointed to by the
blue arrow corresponds to the blue line on the ||-coords (first row left).

The second aspect of good data visualization is the ability of ``interaction'' with the available
figures or plots and ``linking'' among these plots. ``Interaction'' is the ability to select parts
of the data, using GUI actions, that may be of visual interest. Each group should be colored
differently with a transparency level (through an alpha channel) so that different patterns are
distinguishable; this is called brushing. ``Linking'' is the ability to automatically select the
same set of observations on other plots when those observations are selected on one plot. For
example, when the observation represented by the blue line on the ||-coords is selected the
corresponding point should be placed on the map with the same \textit{X} and \textit{Y} coordinates
value on the ||-coords, and the other 7 features (\textit{B1}--\textit{B7}) correspond to its
satellite measures.

Examining the ||-coords plot of this data reveals a weired pattern at the bottom of
\textit{B4}. Selecting this pattern (as brushed in Figure \ref{FigInselberg2008ParCordSolevania1}
(second row left)) surprisingly indicates that those observations are corresponding to the lake of
Slovenia (as brushed in Figure \ref{FigInselberg2008ParCordSolevania1} (second row right)). Thanks
to ``linking''. This is a wonderful shortcut to modeling and clustering this dataset. This visual
inspection gives us the hypothesis that water in this part of the land can be detected from
satellite data by only thresholding the variable \textit{B4}.

For more elaboration on ||-coords, Figure \ref{FigInselbergGeomParCord} \citep[as appears
in][Sec. 10.2.2]{Inselberg2008ParCord} explains the geometry of a 2D line in both perpendicular
coordinates (the usual Cartesian system) and in ||-coords, which may not be intuitive at all for new
comers to ||-coords.

\bigskip

A smart data analyst should study the data visually with many plots and visualization methods than
the ||-coords; e.g., matrix plot, histograms, projection pursuit, among other dozens of available
methods; all should be linked to each other as mentioned above. A snapshot of few of these methods
is illustrated in Figure \ref{FigManyPlots} that is borrowed from \cite{D3gallery}. For a good
reference of the literature of data visualization methods the reader may refer to
\cite{Chen2008DataVisualization}; and for a comprehensive interactive gallery and examples refer to
\cite{D3gallery,mbostock,d3list}. However, we provide this simple example only for illustrating the
concept.

\subsection{Grammar of Graphics (GoG)}\label{sec:elaboration-gog}
A good example to explain the idea of GoG more is adopted from
\cite{Wilkinson2006GrammarOfGraphics}, from where Figures
\ref{FigWilkinsonDeathBirth}--\ref{FigwilkinsonDesignTree} are borrowed. We will not talk here about
the semantics of the figure and the striking information revealed concerning some countries (which
is out of our current scope). We will focus on the GoG that if exists, abstractly and generically
enough, it will produce such a figure and other more complicated figures very efficiently.

The design tree of the GoG of Figure \ref{FigWilkinsonDeathBirth} is drawn in Figure
\ref{FigWilkinsonDeathBirth}. Each line-ending arrow depicts some relation between its two
connectors similar to those adopted in relational databases. The corresponding pseudo code grammar
that describes Figure \ref{FigWilkinsonDeathBirth} is this:
\begin{lstlisting}[style=customphp]
ELEMENT: point(position(birth*death), size(zero), label(country))
ELEMENT: contour(position(smooth.density.kernel.epanechnikov.joint(birth*death)), color.hue())
GUIDE  : form.line(position((0,0),(30,30)), label("Zero Population Growth"))
GUIDE  : axis(dim(1), label("Birth Rate"))
GUIDE  : axis(dim(2), label("Death Rate"))
\end{lstlisting}
Notice that the figure is full of information and many overlaid plots, including many colored
contour plots, 2-D function (the straight line), and plot labels. However, its GoG descriptor is
terse and efficient. Moreover, and most importantly, it is flexible and extensible, which is one of
the most important features of DVP. DVP is designed to provide a scripting language that follows the
GoG of \cite{Wilkinson2006GrammarOfGraphics} to accomplish the extensibility feature discussed in
Section \ref{SecObjectives}.

%%% Local Variables:
%%% mode: latex
%%% TeX-master: "Yousef2016DVP003"
%%% End:

\end{document}